\documentclass[preprint,11pt]{elsarticle}

\usepackage{graphicx}
\usepackage{multirow}
\usepackage{float}
\usepackage{subfigure}
\usepackage{amssymb}
\usepackage{amsmath}

\usepackage{lineno}
\usepackage{verbatim}

\usepackage{geometry}
\usepackage{color}
\usepackage{multicol}
\usepackage{multirow}
\usepackage{booktabs}
\usepackage{array}
\usepackage{threeparttable}
\usepackage{times} 
\usepackage{siunitx}

\journal{}

\makeatletter
\def\ps@pprintTitle{%
\let\@oddhead\@empty
\let\@evenhead\@empty
\def\@oddfoot{}%
\let\@evenfoot\@oddfoot}
\makeatother

\begin{document}

\begin{frontmatter}

\title{Anisotropic thermal characterisation of large-format lithium-ion pouch cells}

\author[label1]{Jie Lin}
\author[label1]{Howie N. Chu}
\author[label1,label2]{Charles W.\ Monroe}
\author[label1,label2]{David A.\ Howey}

\address[label1]{Department of Engineering Science, University of Oxford, Oxford, United Kingdom}
\address[label2]{The Faraday Institution, Harwell Campus, Didcot, United Kingdom}

\begin{abstract}

Temperature strongly impacts battery performance, safety and durability, but modelling heat transfer requires accurately measured thermal properties. Herein we propose new approaches to characterise the heat capacity and anisotropic thermal-conductivity components for lithium-ion pouch cells. Heat capacity was estimated by applying Newton's law of cooling to an insulated container within which the cell was submerged in warmed dielectric fluid. Thermal conductivity was quantified by heating one side of the cell and measuring the opposing temperature distribution with infra-red thermography, then inverse modelling with the anisotropic heat equation. Experiments were performed on commercial 20 Ah lithium iron phosphate (LFP) pouch cells. At 100\% state-of-charge (SOC), the heat capacity of a \SI{489}{g}, \SI{224}{mL} pouch cell was $541$ \si{J.K^{-1}}. The through-plane and in-plane thermal conductivities were respectively $0.52$ and $26.6$ \si{W.m^{-1}.K^{-1}}. Capturing anisotropies in conductivity is important for accurate thermal simulations. State-of-charge dependence was also probed by testing at 50\% SOC: the heat capacity dropped by 6\% and thermal conductivity did not significantly change.

\end{abstract}

\end{frontmatter}

\section{Introduction}
\label{sec:intro} 
Batteries are key enabling devices for the electrification of transport and increased renewable energy generation on the power grid \cite{Gur2018:review,Gibb2021:rise}. Lithium-ion batteries have improved significantly in cost and energy density \cite{ziegler2021re} and are now the standard energy-storage choice in many applications \cite{Chu2016;Path}. Thermal management is an important issue for lithium-ion battery systems, since high temperatures may lead to severe degradation or even catastrophic thermal runaway \cite{Zhu:2019Fast,Finegan:2015In,Feng:2018Review}.

For large-format pouch cells operating at high C-rates, high temperatures and spatial temperature non-uniformities may occur depending on the spatial distribution of electrochemical reactions within the cell \cite{Lin2021:multiscale} and the cooling arrangement. The heat-management challenge has motivated research to develop more thermally conductive electrodes \cite{Koo2017;Toward}, effective thermal management techniques \cite{Amietszajew2019:Hyb,Huang:2020Op,Xu:2020Ne} and improved thermoelectrochemical models \cite{Heinrich2019:Phy,Roder2019:Mod,Deng2018:Safe,Chu:2020pa}. All of these approaches require accurate knowledge of the cell's heat generation and thermophysical properties. Commercial lithium-ion batteries have a multi-layered unit-cell structure, either in a cylindrical or flat-wound jelly roll or a pouch-style stack, with each unit cell containing current collectors, cathode, anode and separator. This complex combination of materials makes ascertaining the overall thermal properties of the cell a difficult challenge.

Calorimetry is a popular method to characterise the heat capacities of batteries. Commercially available accelerating rate calorimeters \cite{bazinski:2015Ex,Bryden2018;Meth,Vertiz2014;The} and differential scanning calorimeters  \cite{Werner:2017Th,Loges2016;Astudy} are commonly used to measure the specific heat of cells and cell materials. These methods are well documented in ASTM Standards, such as ASTM E1981 and E537. Apart from this, `thermal impedance spectroscopy' has also been proposed for cell specific heat capacity measurement \cite{Fleckenstein2013;TIS}. Heat flow calorimeters \cite{Zhang:2014Si} and gradient heat flux sensors \cite{Murashko:2014Th} have also been used to correlate heat flux with temperature change.

Laser-flash and hot-plate methods have been applied for thermal conductivity measurement as standard tests in ASME E1461 and C177, respectively. With these two methods, the commercial xenon flash technique \cite{Maleki:1999Th} and transient planar source technique \cite{Vertiz2014;The} are widely used to characterise the thermal conductivities of prismatic cells or layered cell components. The latter is limited to through-plane thermal conductivity characterisation, while the former can also measure the in-plane thermal conductivity by stacking multiple layers of cell components \cite{Maleki2014;Li}. In addition, photothermal deflection spectroscopy \cite{Werner:2017Th,Loges2016;The} has been used to study the thermal conductivity of different cell components, including electrode coatings, current collectors and separators. Thermoreflectance \cite{Schmidt2008;Pul,Jagannadham2016;The} can be applied to measure thermal conductivity of thin-film electrodes and solid electrolytes, as well as their interfacial conductances. Similar to the transient planar source technique, a constant heat-flux method \cite{Richter2017;The,Zhang:2014Si,Drake2014;Mea} was also established and showed good stability. 

Although there are numerous methods to characterise specific heat and thermal conductivity, some of these may not be feasible for large-format lithium-ion pouch cells. In particular, the large aspect ratio and low thermal conductivity make it difficult to maintain an adiabatic environment or uniform temperature for thermal characterisation, thus hindering the direct application of standard calorimetric or air cooling methods \cite{Sheng2019;An,Zhang2019;Eva}. 
Calculating bulk cell thermophysical properties from mass-averaged cell components and volume-averaged phases could also lead to substantial disagreement with experimental measurements \cite{Vertiz2014;The,Zhang:2014Si}. 

Here we propose novel thermal characterisation approaches for measuring the effective specific heat capacity and anisotropic thermal conductivity of large-format lithium-ion pouch cells, as alternatives to existing techniques (which typically do not give the anisotropic conductivity). We apply the techniques to quantify properties of a commercial 20 Ah LiFePO\(_4\) (LFP)/graphite pouch cell (AMP20M1HD-A, A123 Systems). Heat capacity is characterised using a cell immersed in a dielectric fluid whose temperature differs from the cell, and modelling the transient temperature response of the cell with Newton's law of cooling. The three-dimensional thermal conductivity is obtained by fitting a finite-element thermal model to the surface temperature response triggered by a constant temperature input at the rear cell surface. The dynamic cell temperature evolution was monitored by infra-red thermography. A least-squares optimization algorithm was used to parameterise the thermal-conductivity tensor given the previously measured volumetric heat capacity of the battery.  Validated with aluminium reference samples, these characterisation methods have the merits of simplicity, speed and  accuracy, which is appealing for both scientific research and engineering practice.

\section{Heat-capacity characterisation with transient cooling}
\label{sec:heatcapacity}
\begin{figure}[t]
\centering\includegraphics[width=15cm]{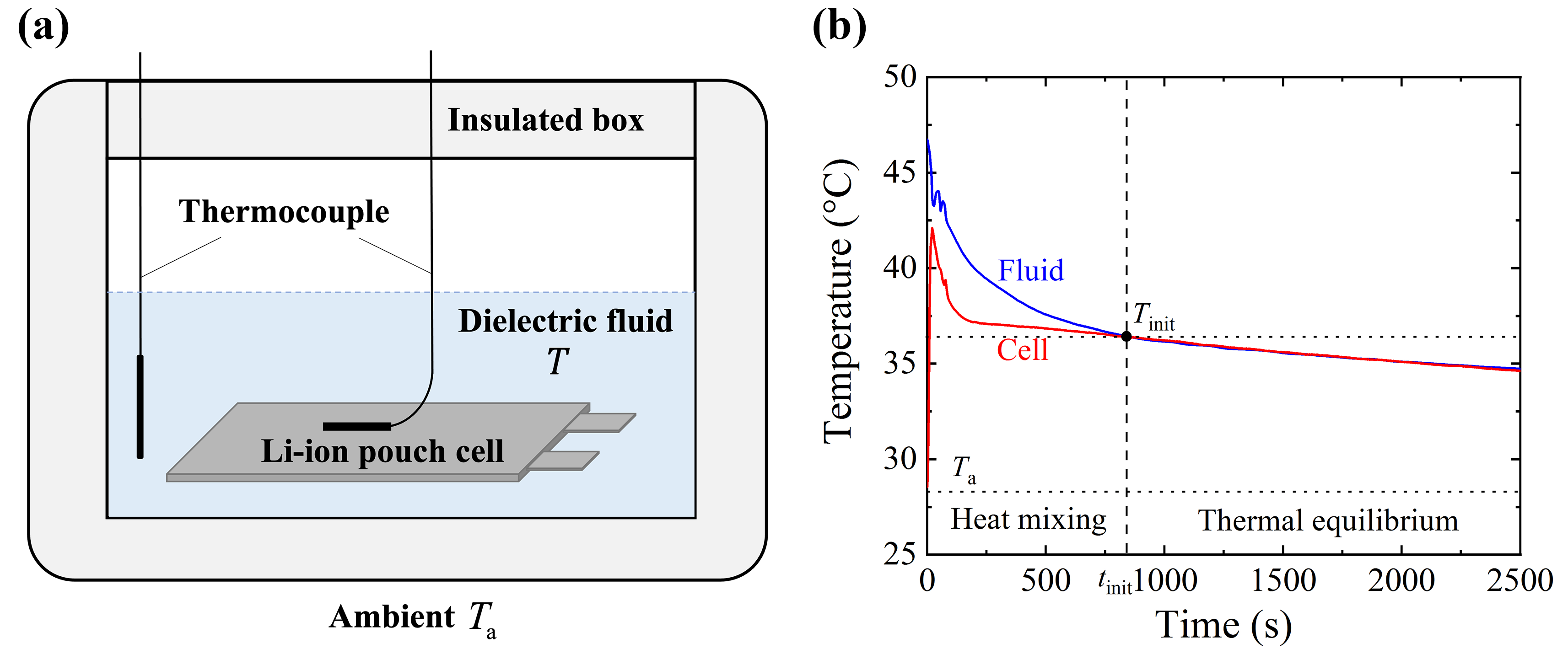}
\caption{Specific heat measurement by transient cooling. (a) Test schematic. (b) Transient temperatures of pouch cell and dielectric fluid, showing the `Heat mixing' regime at times $t < t_{\textrm{init}}$  and the `Thermal equilibrium' regime when the temperatures of the battery and dielectric fluid match, at times $t > t_{\textrm{init}}$. The initial temperature $T_{\textrm{init}}$ that appears in equation \ref{eq:logdecay} is also labeled on the figure.}
\label{fig:cp_apparatus}
\end{figure}

Figure \ref{fig:cp_apparatus}(a) shows the experimental setup for heat capacity measurement.  
A pristine cell at 100\% SOC was placed in an insulated box (Tundra 35, Yeti) at room temperature, which was used as a calorimeter. The walls of the insulated box contain two polyethylene layers ($ca.$ \SI{1}{cm} thick) surrounding a central layer of polyurethane foam (up to \SI{5.0}{cm} thick). Dielectric fluid (Thermal H10, Julabo) heated to a specified elevated temperature in a thermal bath (SC100-A25B, Julabo), was added into the box such that the pouch cell was fully immersed, and the box was sealed to minimize the heat loss to the surroundings. Transient temperature responses of the cell and dielectric fluid were recorded via type T thermocouples (Omega Engineering). The thermocouples were calibrated to \SI{\pm0.1}{\degreeCelsius} absolute accuracy by an ultra precise RTD sensor (1/10 DIN accuracy, Omega Engineering) before use. Prior to the measurement, a reference test was conducted to measure the temperature response of pure dielectric fluid without a sample cell in the calorimeter. 

Since the experimental apparatus is well insulated and the box wall has low thermal conductivity, the system cools predominantly by free convection. Fig.\ \ref{fig:cp_apparatus}(b) shows that after a sufficient relaxation time, thermocouples on the battery surface and in the dielectric fluid achieve equal temperatures. After this time, it is reasonable to assume that the box interior achieves a uniform temperature, even when an immersed cell is present.

Assuming that the box interior has uniform temperature $T$, a general heat balance around the experimental apparatus shown in Fig.\ \ref{fig:cp_apparatus}(a) can be formulated as
\begin{equation}
\label{eq:energybalance}
C_p^\textrm{eff} \frac{dT}{dt} = -\dot{Q},
\end{equation}
in which $C_p^\textrm{eff}$ is the lumped total heat capacity of the system (\si{\joule\per\kelvin}) and $\dot{Q}$ is the heat flow to the surroundings. The heat flow follows Newton's law of cooling,
\begin{equation}
\dot{Q} = hA \left[ T\left( t \right) - T_\textrm{a} \right],
\end{equation}
where $T_\textrm{a}$ is the ambient external temperature, $A$ is the external surface area of the box available for heat transfer, and $h$ is a heat-transfer coefficient intended to account for both conduction through the box wall and free convection around the box. The solution to this problem is
\begin{equation} \label{eq:logdecay}
    \ln \left( \frac{T \left( t \right) - T_\textrm{a}}{T_\textrm{init} - T_\textrm{a}} \right) = - \frac{hAt}{C_p^\textrm{eff}} = - s t.
\end{equation}
Thus the temperature within the box is expected to decay exponentially with respect to time. Here $T_\textrm{init}$ represents the initial temperature after a sufficient equilibration period, represented by the vertical dashed line in Fig.\ \ref{fig:cp_apparatus}, and parameter $s$ indicates the slope of the temperature decay on a semi-log plot.

The heat-capacity measurement technique exploits the general behaviour modelled by Eq.\ \ref{eq:logdecay}, the fact that $h$ and $A$ are essentially identical between tests, and the fact that similar volumes of dielectric fluid are used in every experiment. Thus if one makes a semilog plot of $T - T_\textrm{a}$ for a reference test without the battery, in which the total heat capacity is $C_p^{\textrm{ref}}$, and a semilog plot of $T - T_\textrm{a}$ for a test with the battery, in which case the effective total heat capacity of the box is $C_p^{\textrm{test}}$, the ratio of the slopes of the temperature decay will be
\begin{equation}
    \frac{s^\textrm{ref}}{s^\textrm{test}} = \frac{C_p^{\textrm{test}}}{C_p^{\textrm{ref}}}.
\end{equation}
(Note that the intercept of the linear fit, which depends on $\ln(T_{\textrm{init}}-T_{\textrm{a}})$, is not strictly needed when estimating the slopes, so long as $h$ and $C_p$ are relatively constant.) Finally, the battery's heat capacity can be extracted from this relationship by exploiting the extensivity of total heat capacity, which implies that
\begin{equation}
C_p^{\textrm{battery}} = \left( \frac{m_{\textrm{f}}^{\textrm{ref}} s^{\textrm{ref}} }{m_{\textrm{f}}^{\textrm{test}} s^{\textrm{test}} } - 1 \right) m_{\textrm{f}}^{\textrm{test}} \hat{C}_p^{\textrm{fluid}}
\end{equation}
where $\hat{C}_p^\textrm{fluid}$ is the specific heat capacity of the dielectric fluid (taken to be \SI{1.51}{J.K^{-1}.g^{-1}} here \cite{Dielectricfluid}), $m_\textrm{f}^j$ is the mass of the fluid in experiment $j$, and $C_p^{\textrm{battery}}$ is the total heat capacity of the battery cell. In writing this expression, it has been assumed that both the heat capacity of air in the headspace within the box and that of the box itself are negligible, so that $C_p^\textrm{ref} \approx C_p^\textrm{fluid}$ and $C_p^\textrm{test} 
\approx C_p^\textrm{fluid} + C_p^\textrm{battery}$. Note that this method also requires that the thermal relaxation of the box wall is fast compared to the rate of bulk cooling. 

It should be emphasized that this approach yields the total heat capacity of the battery. Practically it may be more useful to use normalized quantities such as specific heat capacity or volumetric heat capacity. To convert $C_p^\textrm{battery}$ to normalized quantities, the mass of a sample cell was measured to be \SI{489.0}{g} with an analytical balance (IFS 30K0.2DL, Kern \& Sohn GmbH), and the cell's exterior dimensions were measured with calipers as \(205\times155\times7.2\) \si{mm}. The cell volume was also measured to be $V^\textrm{battery} =$ \SI{224.0}{cm^3} by fluid displacement, which agrees with the caliper measurements within 2\%. (Combining the mass and volume measured by liquid displacement yields a bulk cell density of \SI{2.18}{g.cm^{-3}}.)

\section{Anisotropic thermal-conductivity characterisation with thermography}
\label{sec:conductivity}
\begin{figure}[t]
\centering\includegraphics[width=16cm]{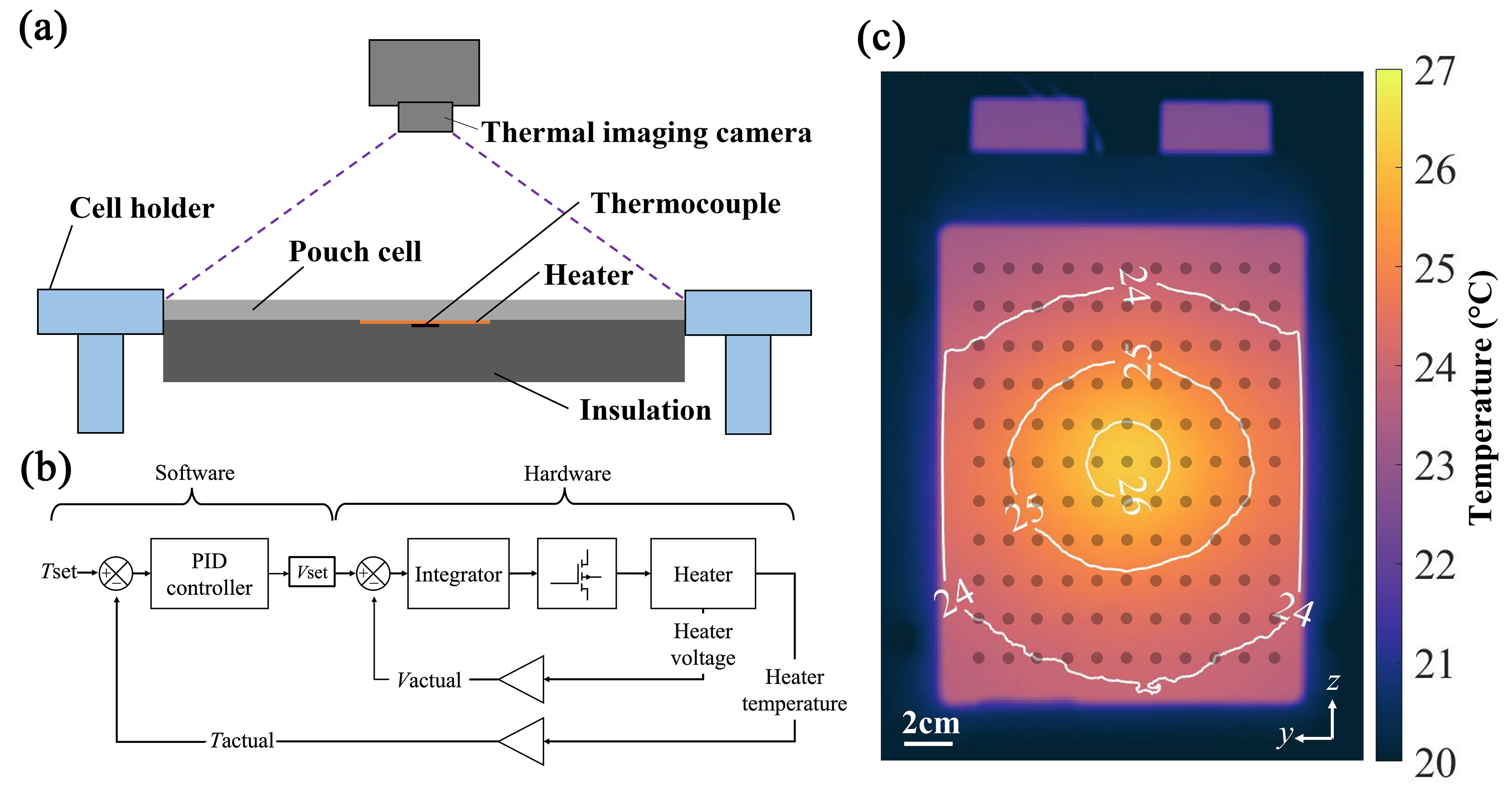}
\caption{Thermal conductivity measurement using a spot-heated cell and infrared thermography. (a) Experimental schematic (side view). (b) Block diagram of control and data acquisition scheme. (c) Thermogram of the battery surface during an experiment, showing  temperature contours in \si{\degreeCelsius} (x-axis into the page). The thermogram also indicates the position of the $11\times11$ grid along which temperature measurements were gathered. The central gridpoint is positioned directly across from the centre of the heater's surface.  
}
\label{fig:cond_measurement}
\end{figure}

Having measured the heat capacity, it becomes possible to measure the effective thermal conductivity within the battery cell by inverse modelling using numerical simulations. Generally the thermal response of the cell follows the heat equation. The form
\begin{equation}
    \frac{C_p^{\textrm{battery}}}{V^\textrm{battery}} \frac{\partial T}{\partial t} = \nabla \cdot \left( \mathbf{k} \cdot \nabla T \right)
\end{equation}
applies when there is no internal heat generation within the battery. Note that this expression involves an effective thermal-conductivity tensor $\mathbf{k}$. Obviously $\mathbf{k}$ is a homogenized quantity that combines the conductivities of current collectors, electrolyte, electrode materials, etc. This homogenization leads one to expect that $\mathbf{k}$ is anisotropic: aligning cartesian coordinates within the cell as depicted in Fig.\ \ref{fig:cond_measurement}(c), the stacked electrode layers are normal to the $x$ axis; consequently one expects the thermal conductivity in the $y$-$z$ plane to differ from the thermal conductivity in the $x$ direction. We assume that the thermal-conductivity tensor is diagonal when the axes are oriented as in Fig.\ \ref{fig:cond_measurement}(c), so that the only non-zero components of $\mathbf{k}$ are $k_{xx}$, $k_{yy}$, and $k_{zz}$, thus
\begin{equation}
\label{eq:3Dthermal}
\frac{C_p^{\textrm{battery}}}{V^\textrm{battery}} \frac{{\partial T}}{{\partial t}} = {k_{xx}}\frac{{{\partial ^2}T}}{{\partial {x^2}}} + {k_{yy}}\frac{{{\partial ^2}T}}{{\partial {y^2}}} + {k_{zz}}\frac{{{\partial ^2}T}}{{\partial {z^2}}}.
\end{equation}
This equation was solved to model an experiment in which a battery was heated from one side using a small disk-shaped heater, embedded in a thermal insulator.

The experimental geometry is shown schematically in Fig.\ \ref{fig:cond_measurement}(a). Pouch cells for testing were laid horizontally on a cell holder. A customized circular polyimide heater (\SI{24}{mm} diameter\(\times\)\SI{0.4}{mm} thickness) was affixed to the geometric centre of the largest rectangular face of the cell, which was oriented downward. A thermocouple  (Type K, Omega Engineering), that had been previously calibrated, was placed at the centre of the heater. This whole back surface was then well insulated with Armaflex insulation material (Class O, \SI{0.02}{W.m^{-1}.K^{-1}}), which was held against the cell surface by tape. Finally, a thermal imaging camera (A655sc, FLIR Systems) was fixed at the top of the test rig to record the surface-temperature response upon heating. Two RTD probes (1/10 DIN accuracy, Omega Engineering) were positioned near the cell holder to record the ambient temperature and were used as a reference for the camera. The heater was connected to a bespoke control circuit following the method of Howey et al.\ \cite{howey2010radially}, which regulated power to ensure a constant surface temperature. Taking the thermocouple reading as the feedback signal, a proportional–integral–derivative (PID) control program was executed in Labview to fix a constant thermocouple temperature as per the block diagram in Fig.\ \ref{fig:cond_measurement}(b)). The temperature setpoint was generally reached on a very short timescale (order seconds). 

To model this configuration, energy-balance equation \ref{eq:3Dthermal} was simulated with a constant-temperature boundary condition (Dirichlet type) where the heater contacted the back of the battery cell; an adiabatic (Neumann type) boundary condition on the rest of the back; and Newton's law of cooling (Robin type) on all the other exterior surfaces of the cell. Mathematically,
\begin{equation}
\begin{array}{cl}
T=T_\textrm{h} & (\textrm{Dirichlet}), \\
-\vec{n} \cdot \mathbf{k} \cdot \nabla T =0 & (\textrm{Neumann}) , \\
- \vec{n} \cdot \mathbf{k} \cdot \nabla T =h(T-T_\textrm{a}) & (\textrm{Robin}), 
\end{array}
\label{eq:boundtype}
\end{equation}
where $T_\textrm{h}$ is the heater's surface temperature, $h$ is a heat-transfer coefficient ostensibly describing free convection, and $\vec{n}$ is an outward surface normal vector; $\mathbf{k}$ represents the diagonal thermal-conductivity tensor.

The thermal model was solved numerically with COMSOL Multiphysics software, assuming that the battery was initially at ambient temperature, and using the actual $T_\textrm{h}$ values recorded by the heater's controller as a transient Dirichlet condition. Source code is available on GitHub \cite{Github2021}. These simulations produce the transient surface-temperature distribution on the surface of the cell opposite the heater as functions of the ambient temperature, heat-transfer coefficient, volumetric battery heat capacity, and the three thermal-conductivity tensor components.

To fit the thermal conductivity, a uniform \(11\times 11\) matrix of surface temperature points was defined around the heater centre, evenly spaced within the dash-dotted rectangular region shown in Fig.\ \ref{fig:cond_measurement}(d), for a total of 121 points across the cell surface. To quantify the discrepancy between simulation and experiment, an objective function $f$ is defined as

\begin{equation}
\begin{aligned}
\label{eq:costfunc}
f=\sum_{i=1}^{11} \sum_{j=1}^{11} \sum_{k=1}^{N}\left ( \frac{T_{ijk}^\textrm{sim}-T_{ijk}^\textrm{exp}}{T_{ijk}^\textrm{exp}} \right )^{2}
\end{aligned}
\end{equation}
\noindent where indices $i,j$ designate locations on the surface grid and $k$ is the time step; $N$ is the number of data sets in the time series. During fitting, the components of thermal conductivity and the heat-transfer coefficient were adjusted to minimize \(f\) via a nonlinear least-squares optimization algorithm.

\section{Results and Discussion}
\label{sec:results}

\subsection{Heat capacity}
\begin{figure}
\centering\includegraphics[width=12cm]{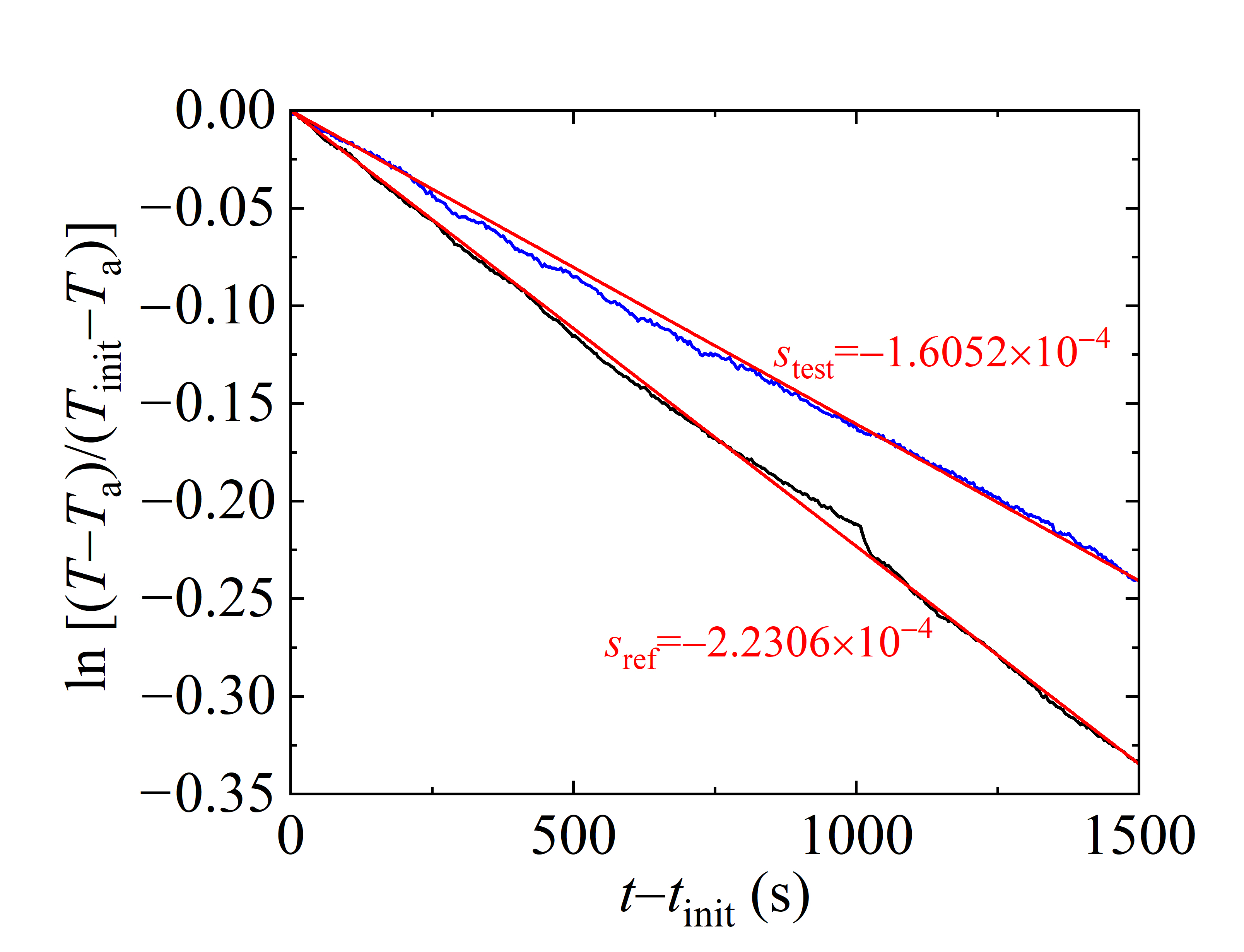}
\caption{Run \#1 of specific-heat measurement by transient cooling for a pouch cell, showing the temperature responses during reference (black) and test steps (blue). Linear fits for each test are shown in red.}
\label{fig:results_cp}
\end{figure}

Heat-capacity measurements were always performed in pairs; for each run, a reference step with dielectric fluid alone was carried out prior to each test step with a sample. The empty chamber was typically rested for more than an hour between pairs of runs. 
For both reference and test steps, the dielectric fluid was initially heated to \emph{ca.} \SI{45}{\degreeCelsius} before being transfjerred to the insulated box, after which its tare weight was recorded. The pouch cell was always at ambient temperature (ca.\ 25-28 \si{\degreeCelsius}) before being immersed in the fluid; test data were only processed after the battery-surface and fluid temperatures equilibrated (cf.\ Fig. \ref{fig:cp_apparatus}). Figure \ref{fig:results_cp} illustrates the temperature responses of the apparatus during the reference and tests steps of Run 1 as an example. Reference and test data during the thermal equilibrium period established during the test step (cf.\ Figure \ref{fig:cp_apparatus}(b)) are plotted on a semi-log scale. Linear fits are provided for comparison, confirming the expectations that the thermal relaxation is linear in the semi-log representation, and also that temperature in the test step decays with a shallower slope. Four runs of the experiment were performed, yielding the heat capacity measurements in Table \ref{tab:heatcap_res}. The heat capacity of the cell was \SI{541}{J.K^{-1}}; standard error of the mean across the four runs was \SI{13}{J.K^{-1}}, approximately $2.4$\%. 

To validate the method, an aluminium alloy (Grade 5251-H22) plate with similar dimensions to the pouch cell (\(205\times161\times6\) \si{mm}) was also studied with the transient cooling apparatus. The density, specific heat and thermal conductivity of the aluminium alloy have been reported previously as \SI{2690}{kg.m^{-3}}, \SI{900}{J.kg^{-1}.K^{-1}} and \SI{149}{W.m^{-1}.K^{-1}}, respectively \cite{aluminiumdata,aluminiumdatabase}. Table \ref{tab:heatcap_res} also presents data from four transient-cooling runs using the aluminium plate as a test sample instead of the battery.

\begin{table}
\centering
\caption{Specific heat of the aluminium reference sample and LFP pouch cell @100\% SOC.}
\label{tab:heatcap_res}
\begin{threeparttable}
\begin{tabular}{c|c|cc|cc|cc} 
\hline
\multirow{2}{*}{Sample}    & \multirow{2}{*}{Run} & \multicolumn{2}{c|}{Reference$^*$}  & \multicolumn{2}{c|}{Test$^{**}$} & \multicolumn{2}{c}{Heat capacity}    \\
 &   & \(m_\textrm{f}^\textrm{ref}\) [\si{g}]    & \(s^\textrm{ref}~[10^{-4}\;\si{s^{-1}}] \) & \(m_\textrm{f}^\textrm{test}\) [\si{g}] & \(s^\textrm{test}~[10^{-4}\;\si{s^{-1}}] \) & \(C_p\) [\si{J.K^{-1}}] &  \(\hat{C}_p\) [\si{J.g^{-1}.K^{-1}}] \\ 
\hline
\multirow{4}{*}{Al plate} & 1    & $969.6$  & $2.7225$  & $1001.6$    & $2.0063$   &  $474.2$  &  $0.8890$  \\
 & 2 & $1027.1$   & $2.6461$   & $917.1$   & $2.2031$ & $478.0$  & $0.8961$  \\
 & 3   & $891.6$ & $2.8467$   & $937.6$  & $1.9991$ & $501.4$   &  $0.9400$  \\
 & 4     & $948.7$     & $2.8003$    & $936.4$   & $2.1132$ & $484.4$ &  $0.9081$  \\ 
\hline
\multirow{4}{*}{Battery}   & 1    & $990.4$ & $2.2306$  & $1000.8$ & $1.6052$ & $567.0$ &  $1.1596$     \\
 & 2  & $1021.1$  & $2.3034$  & $989.4$   & $1.7756$ & $506.2$ & $1.0352$   \\
 & 3  & $992.3$   & $2.1798$  & $1003.2$  & $1.5797$ & $552.7$ & $1.1305$    \\
 & 4  & $1013.4$  & $2.2198$  & $991.5$   & $1.6702$ & $536.6$ & $1.0972$ \\
\hline
\end{tabular}
\begin{tablenotes}
\footnotesize
\item[*]  The masses of the Al plate and battery are \SI{533.4}{g} and \SI{489.0}{g}, respectively.
\item[**]  The specific heat of the dielectric fluid is \SI{1.510}{J.g^{-1}.K^{-1}} at 20-40 \si{\degreeCelsius}, provided by the manufacturer.
\end{tablenotes}
\end{threeparttable}
\end{table}%

The average specific heat of the Al plate was found to be \SI{0.908}{J.g^{-1}.K^{-1}}, with standard error of the mean of \SI{0.011}{J.g^{-1}.K^{-1}}, or 1.2\%. Note that this standard error, despite being small, is greater than the 1\% deviation of the measured heat capacity from the literature value for aluminium. Thus our best estimate of heat capacity for aluminium agrees with the literature value within the experimental error. Runs with the LFP pouch cell at 100\% SOC yield an average heat capacity of \SI{541}{J.K^{-1}}, with a standard error of the mean of \SI{13}{J.K^{-1}} (2.4\%). 

Note that runs were also conducted with the cell at 50\% SOC;  these results are provided in Table S1 of the supplementary information. The cell's heat capacity at 50\% SOC was \SI{507}{J.K^{-1}} with standard error of 1.8\%, i.e. \SI{9}{J.K^{-1}}. These data are suggestive, but more experiments would be required to conclude decisively whether the pouch cell's heat capacity varies significantly with SOC.

Previous studies have reported specific heat capacities of LFP cells, which are compared to the present results in Table \ref{tab:cond_res_lit}. These values range from $0.95$ to \SI{1.70}{J.g^{-1}.K^{-1}}; the majority are near \SI{1}{J.g^{-1}.K^{-1}}. Given that the specific heat capacity at 100\% SOC is $1.11\pm0.05$ \si{J.g^{-1}.K^{-1}}, the present method yields results in good agreement with prior observations. 

\begin{table}
\centering
\caption{Specific heat capacities reported for LFP/graphite cells.}
\label{tab:heatcap_lit}
\resizebox{\linewidth}{!}{%
\begin{tabular}{llclc} 
\hline
\multicolumn{1}{c}{Authors} & \multicolumn{1}{c}{Cell geometry} & Capacity (\si{Ah}) & \multicolumn{1}{c}{Method}     & \(\hat{C}_p\) (\si{J.g^{-1}.K^{-1}})  \\ 
\hline
Prada et al. \cite{Prada2012;Sim}    & Cylindrical       & $2.3$     & Accelerating rate calorimetry         & $1.100$         \\
Fleckenstein et al. \cite{Fleckenstein2013;TIS}  & Cylindrical    & $4.4$   & Thermal impedance spectroscopy & $0.958$          \\
Bazinski et al. \cite{bazinski:2015Ex}      & Pouch        & $14$    & Accelerating rate calorimetry       & $1.10$-$1.68$    \\
Bryden et al. \cite{Bryden2018;Meth}       & Cylindrical   & $2.5$      & Internal temperature sensors   & $1.169$         \\
Sheng et al. \cite{Sheng2019;An}       & Prismatic    & $8$    & Improved calorimetry          & $1.08$-$1.27$    \\
Chu et al. \cite{Chu:2020pa}       & Pouch (various SOC)       & $20$     & Lock-in thermography       & $0.93$-$1.06$     \\
This work       & Pouch (100\% SOC)         & $20$     & Transient cooling       & $1.10$     \\
\hline
\end{tabular}
}
\end{table}

\subsection{Thermal conductivity}
Thermal conductivity measurements were made by recording the transient thermography response of the cell surface to a step change in heater temperature. At the beginning of each run, the  heater was stepped from ambient temperature up to  \SI{50}{\degreeCelsius}. Thermograms of the cell surface, as well as heater voltage, current and temperature, were recorded at \SI{1}{s} intervals. Each test had a total duration of  \SI{900}{s}. The aluminium reference sample was used to validate the thermal conductivity measurement approach. Table \ref{tab:alu_cond} presents the estimated thermal conductivity \(k\) from each of these tests, gathered under the assumption that thermal conductivity is isotropic ($k_{xx} = k_{yy} = k_{zz}$ in $\mathbf{k}$). The mean is \SI{144}{W.m^{-1}.K^{-1}} and the standard deviation of the mean is \SI{1.3}{W.m^{-1}.K^{-1}} (0.9\%). Similarly the mean estimated heat-transfer coefficient is \SI{15.4}{W.m^{-2}.K^{-1}}, with a standard deviation of the mean of \SI{0.35}{W.m^{-2}.K^{-1}} (2.3\%). These numbers reasonably agree with literature for aluminium alloy 5251 ($k = 149$ \si{W.m^{-1}.K^{-1}} \cite{aluminiumdata}) and heat transfer for free convection from a flat plate ($h=2$--$25$ \si{W.m^{-2}.K^{-1}} \cite{Incropera2011}).

\begin{table}
\centering
\caption{Thermal conductivity of the Al plate.}
\label{tab:alu_cond}
\begin{threeparttable}
\begin{tabular}{c|cc} 
\hline
Test number~ & \(k\) (\si{W.m^{-1}.K^{-1}})& \(h\) (\si{W.m^{-2}.K^{-1}})   \\
\hline
1                             & $141$      & $15.2$          \\
2                             & $143$      & $14.7$          \\
3                             & $147$      & $15.4$			\\
4                             & $145$      & $16.4$          \\
\hline
\end{tabular}
\end{threeparttable}
\end{table}

\begin{figure}
\centering\includegraphics[width=16cm]{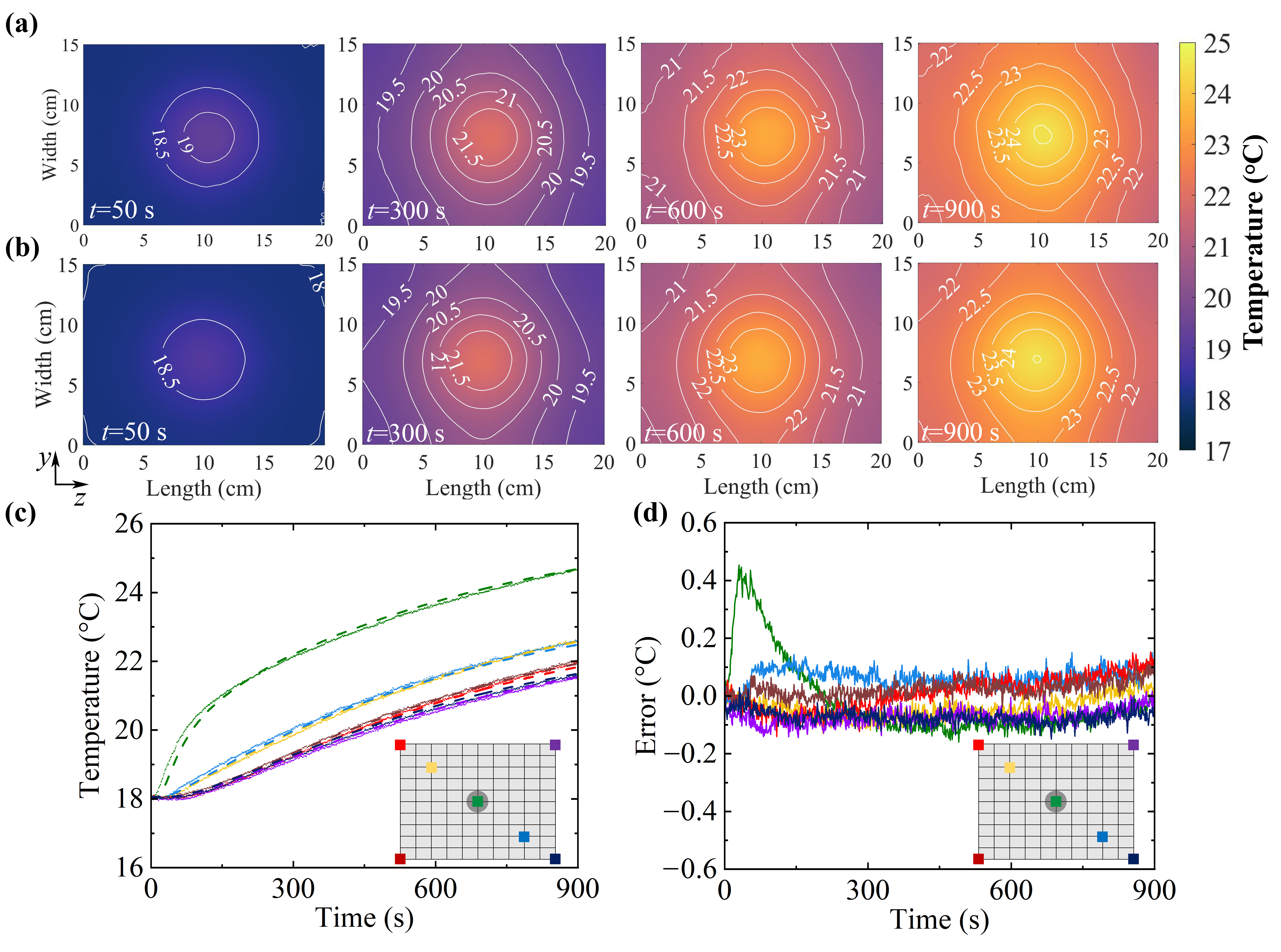}
\caption{Test \#1 of the thermal conductivity characterization. (a) Thermal images of cell front surface at $t=$ \SI{50}{s}, \SI{300}{s}, \SI{600}{s} and \SI{900}{s}. (b) Simulation output  with best-fit parameters based on the experiment shown in panel (a); note that the tabs are at the right edge of the images. Measured (points) and simulated (dashed lines) surface-temperature data (c) and experimental error (d) at selected locations colour-coded in the insets.  
}
\label{fig:cond_results}
\end{figure}

Next, identical thermography experiments were conducted on the lithium-ion pouch cell at 100\% SOC. Figure \ref{fig:cond_results}(a) shows the typical evolution of temperature on the exposed cell surface yielded by thermography, and Fig.\ \ref{fig:cond_results}(b) shows the best-fit model results for comparison. Figure \ref{fig:cond_results}(c) presents plots of the transient temperatures at various points across the cell surface. As expected, the position on the exposed surface directly opposite the centre of the heating element is always hottest. Its temperature does not begin to rise until about \SI{50}{s} after the heater is switched on, because some time is taken for the thermal boundary layer to travel through the pouch. This lag time depends directly on the through-plane thermal conductivity component $k_{xx}$. Notably, the shapes of the temperature contours at \SI{50}{s} are reasonably circular, indicating that the thermal-conductivity components in-plane (in the $y$ and $z$ directions) are nearly equal. At longer times, the temperature rises across the entire cell surface. Rather than exhibiting boundary-layer behaviour, the point directly opposite the heater centre begins to relax exponentially upwards, towards a steady-state temperature. During this time, the temperature profile also spreads beyond the heater's position due to in-plane heat conduction. The contours deviate from circularity because of edge effects. Notably, the lag time after which temperature begins to rise at a given point on the surface increases with respect to the point's distance from the centre of the heater, because the thermal boundary layer must travel further to reach it. Finally, observe that the temperature on the side near the tabs (the right of the thermograms in Fig.\ \ref{fig:cond_results}) differs slightly from that on the left. This owes to direct conduction of heat through the tabs, which slightly alters the thermal boundary conditions. 

Table \ref{tab:cond_results} summarises the results of four thermal-conductivity measurements. The mean through-plane thermal conductivity $k_{xx}$ was \SI{0.51}{W.m^{-1}.K^{-1}}, with a standard error of $1.8\%$. This is about 50 times smaller than the in-plane thermal conductivities, $k_{yy}=\:$\SI{26.45}{W.m^{-1}.K^{-1}}$\pm 1.5\%$ and $k_{zz}=\:$\SI{26.75}{W.m^{-1}.K^{-1}}$\pm 1.7\%$. Notably the in-plane conductivity components agree within the experimental error, but the through-plane conductivity component differs significantly. Therefore the in-plane conductivity $k_{\parallel}$ will be treated as a single quantity, $k_{yy}=k_{zz}=k_{\parallel}$ henceforth. Under this assumption the experimental in-plane conductivity was found to be $k_{\parallel}=$\SI{26.6}{W.m^{-1}.K^{-1}}$\pm 1.1\%$.

To probe the SOC dependence of thermal conductivity, additional measurements were carried out at 50\% SOC. Results of these experiments are presented in Table S2 of the Supplementary Information. The average \(k_{xx}\), \(k_{yy}\) and \(k_{zz}\) did not change beyond the bounds of experimental error.

\begin{table}
\centering
\caption{Thermal conductivity components of the LFP pouch cell at 100\% SOC.}
\label{tab:cond_results}
\begin{tabular}{c|cccc} 
\hline
Test number~ & \(k_{xx}\) (\si{W.m^{-1}.K^{-1}}) &\( k_{yy} \) (\si{W.m^{-1}.K^{-1}}))&\( k_{zz} \) (\si{W.m^{-1}.K^{-1}})& \(h\)  (\si{W.m^{-2}.K^{-1}}) \\
\hline
1                             & $0.513$      & $26.6$      & $27.0$      & $18.5$                           \\
2                             & $0.503$      & $25.9$      & $26.2$      & $18.3$                          \\
3                             & $0.503$       &$25.8$       & $25.9$     & $17.7$                        \\
4                             & $0.544$      & $27.5$      & $27.9$      & $17.2$                          \\
\hline
\end{tabular}
\end{table}%

The through-plane and in-plane thermal conductivity measured in this work are compared with previous studies in Table \ref{tab:cond_res_lit}, showing fair agreement. 

\begin{table}
\centering
\caption{Thermal conductivity of LFP/graphite cells reported in the literature.}
\label{tab:cond_res_lit}
\begin{threeparttable}
\resizebox{\linewidth}{!}{%
\begin{tabular}{llclcc} 
\hline
\multicolumn{1}{c}{Authors} & \multicolumn{1}{c}{Cell geometry} & Capacity (Ah) & \multicolumn{1}{c}{Method}     & \(k_{\bot}\) (W/(m\(\cdot\)K))          & \(k_{\parallel}\) (W/(m\(\cdot\)K))          \\ 
\hline
Bazinski et al. \cite{bazinski:2015Ex}            & Pouch                             & $14$            & Constant heat flux             & $0.34-0.37$   & --          \\
Vertiz et al. \cite{Vertiz2014;The}             & Pouch                             & $14$            & Hot plate                      & $0.235-0.284$ & --          \\
Fleckenstein et al. \cite{Fleckenstein2013;TIS}       & Cylindrical                       & $4.4$           & Thermal impedance spectroscopy & $0.35$        & --          \\
Drake et al. \cite{Drake2014;Mea}                & Cylindrical                       & --             & Constant heat flux             & $0.15-0.20$   & $30.4-32.0$  \\
This work              & Pouch                       & $20$             & Spot temperature-step transient             & $0.51$   & $26.6$  \\
\hline
\end{tabular}
}
\begin{tablenotes}
\footnotesize
\item[*] Subscripts "\(\bot\)" and "\(\parallel\)" denote the through-plane and in-plane directions, respectively.
\end{tablenotes}
\end{threeparttable}
\end{table}

\section{Conclusion}
\label{sec:conclusion}
Novel methods were proposed and implemented to quantify heat capacity and three-dimensional thermal-conductivity components for large-format lithium-ion pouch cells. Heat capacity was measured by tracking the thermal relaxation of a diabatic calorimetry cell that enclosed a battery immersed in a hot working fluid. Newton's law of cooling was exploited to determine the heat capacity from the different relaxation times during reference and test runs. Subsequently, the anisotropic thermal conductivity components were estimated by fitting a 3D finite-elment model to the transient surface-temperature profiles imposed by a disc-shaped, isothermal heater in contact with the rear of the pouch cell. The front-surface temperature was monitored transiently by infrared thermography as the heater underwent a step change in temperature. In both cases, the accuracy of the technique was confirmed by comparing measurements of battery cells to measurements of an aluminium plate, for which all relevant properties were known. 

Accurate values for a battery's heat capacity and thermal conductivity are crucial to thermal modelling, and are not widely reported. The characterisation methods presented here are simple to execute and could be useful to researchers who require values of thermal parameters. Experiments quantifying heat capacity and thermal-conductivity components were found to have good precision (within 1-3\% across all experiments), as well as comparing well with the available literature data for LFP cells. 
Notably, anisotropic thermal-conductivity estimates are not widely available in the battery literature. We find that considering this anisotropy is critical because the in-plane and through-plane conductivities differ by well over an order of magnitude; this has significant impact on thermal transients that should be considered within battery models. The state-of-charge dependence of thermal properties could also be worth considering in the future, but preliminary data here suggest that its importance is secondary. 

\section*{Acknowledgements}
The authors gratefully acknowledge funding from the EPSRC Translational Energy Storage Diagnostics (TRENDs) project (EP/R020973/1) and the STFC Futures Early Career Award, as well as the Faraday Institution Multiscale Modelling Project (subaward FIRG003 under grant EP/P003532/1). We are grateful to Peter Get from Oxford Robotics Institute for constructing and testing the circuit used to drive the heater.

\bibliographystyle{elsarticle-num-names}
\bibliography{Reference.bib}



\end{document}